\documentclass[letter]{aa} 

\usepackage{graphicx, comment}
\usepackage{txfonts}
\usepackage[colorlinks=true]{hyperref}
\newcommand{\C}{3C\,84}

\newcommand{\Bzero}{1.80-4.0}

\newcommand{\Bone}{60-180}

\newcommand{\Bfromjetpower}{30-60}%

\newcommand{\OpenAngle}{2.8${}^{\circ}$-20${}^{\circ}$} 
\newcommand{\OpenAnglealt}{3^{\circ}-20^{\circ}}

\newcommand{\Dfrange}{1.18-1.25}

\newcommand{\bapp}{0.1-0.2\,c}

\newcommand{\Omegarnualt}{3.10\pm0.13}
\newcommand{\Omegarnuone}{2.08\,$\pm$\,0.15}
\newcommand{\Omegarnutwo}{3.69\,$\pm$\,0.26}
\newcommand{\Omegarnurange}{1.93-3.95}

\newcommand{\coreshift}{76-90}

\newcommand{\imageff}{19}
\newcommand{\imagefe}{15}

\newcommand{\kr}{1}
\newcommand{\kralt}{1}
\newcommand{\propconst}{83\,$\pm$\,7} 
\newcommand{\propconstalt}{83\,\pm\,7}%

\newcommand{\bindex}{1}
\newcommand{\bindexalt}{1}

\newcommand{\bindexthreealt}{\approx 1.7}

\newcommand{\PA}{$-20^{\circ}\pm14^{\circ}$}

\newcommand{\DRAfifteen}{60\,$\pm$\,58}
\newcommand{\DDECfifteen}{240\,$\pm$\,53}
\newcommand{\DRAfourtythree}{20\,$\pm$\,52}
\newcommand{\DDECfourtythree}{120\,$\pm$\,46}
\newcommand{\RAuncertaintyff}{30}
\newcommand{\DECuncertaintyff}{30}
\newcommand{\RAuncertaintyfe}{40}
\newcommand{\DECuncertaintyfe}{20}

\newcommand{\Absff}{$369\pm80$}
\newcommand{\Absef}{$122\pm72$}

\newcommand{\Dpc}{0.028-0.11}
\newcommand{\DRs}{400-1500}

\newcommand{\Projpc}{0.030\,$\pm$\,0.002}
\newcommand{\ProjRs}{413\,$\pm$\,34}

\newcommand{\cff}{0.9975}

\newcommand{\cfe}{0.9995}

\begin{document}

   \title{Pinpointing the jet apex of \C}


   \author{G. F. Paraschos\inst{1}, J.-Y. Kim\inst{2,1}, T. P. Krichbaum\inst{1}, J. A. Zensus\inst{1}
          }
   \authorrunning{G. F. Paraschos et al.}
   \institute{$^{1}$Max-Planck-Institut f\"ur Radioastronomie, Auf dem H\"ugel 69, Bonn\\ 
              $^{}$\ \email{gfparaschos@mpifr-bonn.mpg.de}\\
              $^{2}$Korea Astronomy and Space Science Institute, 776 Daedeokdae-ro, Yuseong-gu, Daejeon, 30455, Korea \\
             }

   \date{Received 10/03/2021; accepted 09/06/2021}

 
  \abstract{
Nearby radio galaxies that contain jets are extensively studied with very long baseline interferometry (VLBI), 
  addressing jet launching and the physical mechanisms at play around massive black holes. \C\ is unique in this regard because the combination of its proximity and large super massive black hole (SMBH) mass provides a high spatial resolution to resolve the complex structure at the jet base. 
  For \C,\ an angular scale of 50\,$\mu$as corresponds to $200-250$\, Schwarzschild radii (R$_{\text{s}}$).
  Recent {\it RadioAstron} VLBI imaging at 22\,GHz has revealed an east-west elongated feature at the northern end of the VLBI jet, which challenges past interpretations. Here we propose instead that the jet apex is not located within the 22\,GHz VLBI core region but 
  more upstream in the jet.
  We base our arguments on a 2D cross-correlation analysis of quasi-simultaneously obtained VLBI images at 15, 43, and 86\,GHz, which measures the opacity shift of the VLBI core in \C. 
  With the assumption of the power-law index ($k_r$) of the core shift being set to 1, 
  we find the jet apex to be located $\propconstalt$\,$\mu$as north (upstream) of the 86\,GHz VLBI core. 
  Depending on the assumptions for $k_r$ and the particle number density power-law index, $n$, we find a mixed toroidal-poloidal magnetic field configuration, consistent with a region that is offset from the central engine by about \DRs\,R$_s$.
  The measured core shift is then used to estimate the magnetic field strength, which amounts to B=\Bzero\,G near the 86\,GHz VLBI core. We discuss some physical implications of these findings.}

  %



   \maketitle
%

\section{Introduction}

\C\ is a peculiar Seyfert 1.5-type radio galaxy \citep{2006A&A...455..773V}. It is located relatively nearby, at a distance of 76.9\,Mpc ($z$\,= 0.0176) \citep{1992ApJS...83...29S} \footnote{We assume $\Lambda$ cold dark matter cosmology with H$_{0}$\,=\,71\,km/s/Mpc, $\Omega_\Lambda$\,=\,0.73, and $\Omega_M$\,=\,0.27.}, and harbours a central super massive black hole (SMBH) of $M_{BH} \sim 9 \times 10^8 M_{\odot}$ \citep{2013MNRAS.429.2315S}. This makes \C\ a prime target to pinpoint the location of the SMBH and to study the magnetic field in the very inner jet region for this enigmatic radio galaxy.
\C\ has been studied with centimetre- and millimetre-very long baseline interferometry (VLBI) for decades (e.g. by \citealt{1994ApJ...430L..45W}, \citealt{1998ApJ...498L.111D}, \citealt{2000ApJ...530..233W}, \citealt{2012ApJ...746..140S}, \citealt{2014ApJ...785...53N}, \citealt{2018NatAs...2..472G}, and \citealt{2019A&A...622A.196K}, among others). It features a complex two-sided jet \citep{1994ApJ...430L..41V, 1994ApJ...430L..45W, 2017MNRAS.465L..94F, 2020ApJ...895...35W}, which commonly exhibits moving radio emitting features (blobs) that accelerate with apparent speeds from $\leq0.1c$ 
on sub-milliarcsecond (sub-mas) scales to $0.5c$ on milliarcsecond (mas) scales \citep{1992A&A...260...33K, 1998ApJ...498L.111D, 2021ApJ...911...19P}. 
Bright and fast moving knots have been tracked over the years \citep{1990ApJ...360L..43D, 1998ApJ...498L.111D}, and two components have also been ejected in the southern jet, called C2 and C3 (following the naming convention of \citealt{2014ApJ...785...53N}). Recent high-resolution VLBI imaging of \C\ with the RadioAstron space telescope has revealed a limb-brightened double-railed jet, possibly anchored in a very wide jet base of $\sim$250\,R$_s$ diameter \citep{2018NatAs...2..472G}. This raises the question of whether the jet in \C\ is launched via magneto-centrifugal acceleration from the accretion disk -- the \citealt{1982MNRAS.199..883B} (BP) model -- or via 
energy extraction 
directly from the ergosphere of the spinning central black hole -- the \citealt{1977MNRAS.179..433B} (BZ) model.

The unknown location of the SMBH in \C\ can be estimated by assuming its proximity to the
jet apex. The latter can be determined from high-resolution VLBI imaging in the millimetre bands. 
VLBI imaging at these short wavelengths not only provides a higher angular resolution than in the centimetre bands, but also helps to overcome opacity effects; these effects are caused by the 
synchrotron self-absorption in the jet and by free-free absorption from the circum-nuclear gas of the accretion flow, which partially obscures the counter-jet and jet base \citep{1994ApJ...430L..45W, 2017MNRAS.465L..94F, 2019A&A...622A.196K}.

Here we present a new study of \C, which measures the VLBI core shift. 
Such core shifts are also observed in the jets of several other galaxies and blazars
(e.g. \citealt{1998A&A...330...79L, 2013A&A...557A.105F, 2011Natur.477..185H, 2012A&A...545A.113P, 2013EPJWC..6108004H, 2021ApJ...909...76P}).
Active galactic nucleus (AGN) jets emit synchrotron radiation, which is susceptible to synchrotron self-absorption.
Synchrotron self-absorption is frequency dependent \citep{1979rpa..book.....R}. When the
VLBI core is associated with the $\tau=1$ absorption surface
(e.g. \citealt{1981ApJ...243..700K, 1998A&A...330...79L}), its position becomes frequency dependent. 
The optically thick VLBI core region becomes more transparent at higher frequencies and shifts in the direction of the opacity gradient. For a conical, homogeneous, and straight jet, this opacity gradient points towards the jet apex. With this assumption, the observed position shift of the VLBI core can be used to estimate the magnitude and topology of the magnetic field at the jet base (i.e. toroidal vs. poloidal field; see \citealt{1998A&AS..132..261L, 2005ApJ...619...73H, 2006Natur.440...58V, 2009MNRAS.400...26O, 2012A&A...545A.113P, 2013A&A...557A.105F}).

This paper is organised as follows: In Sect. \ref{sec:Results} we briefly summarise the observations and data reduction and then present our spectral and core shift analysis results. We discuss our results in Sect. \ref{sec:Disc} and present our conclusions in Sect. \ref{sec:Conclusions}.

\section{Data, analysis, and results}\label{sec:Results}

In this section we present the 2D cross-correlation analysis used to produce the spectral index maps and core shift. We define the spectral index $\alpha$ as $S\propto\nu^{+\alpha}$. The observations were made with the Very Long Baseline Array (VLBA) at 15 and 43\,GHz and the Global Millimeter VLBI Array (GMVA) at 86\,GHz during the period of 11-18 May, 2015. Total intensity and polarisation imaging results have already been published; for the details we refer to \cite{2019A&A...622A.196K}. In this follow-up paper we used the same data but now focus on the spectral properties and opacity shift in the VLBI core region.

\begin{figure*}
   \centering
   \includegraphics[scale=0.25]{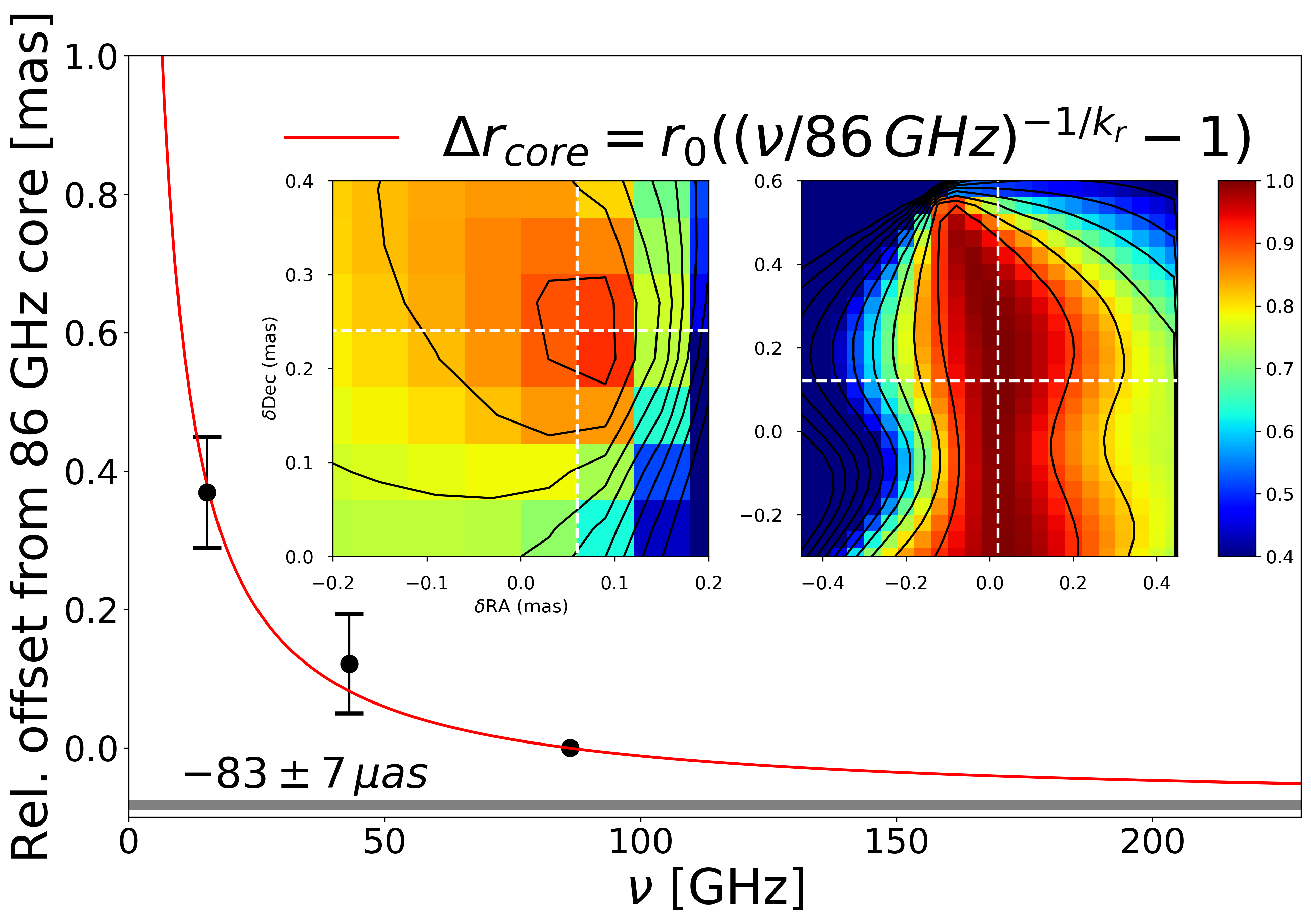}
   \includegraphics[scale=0.25]{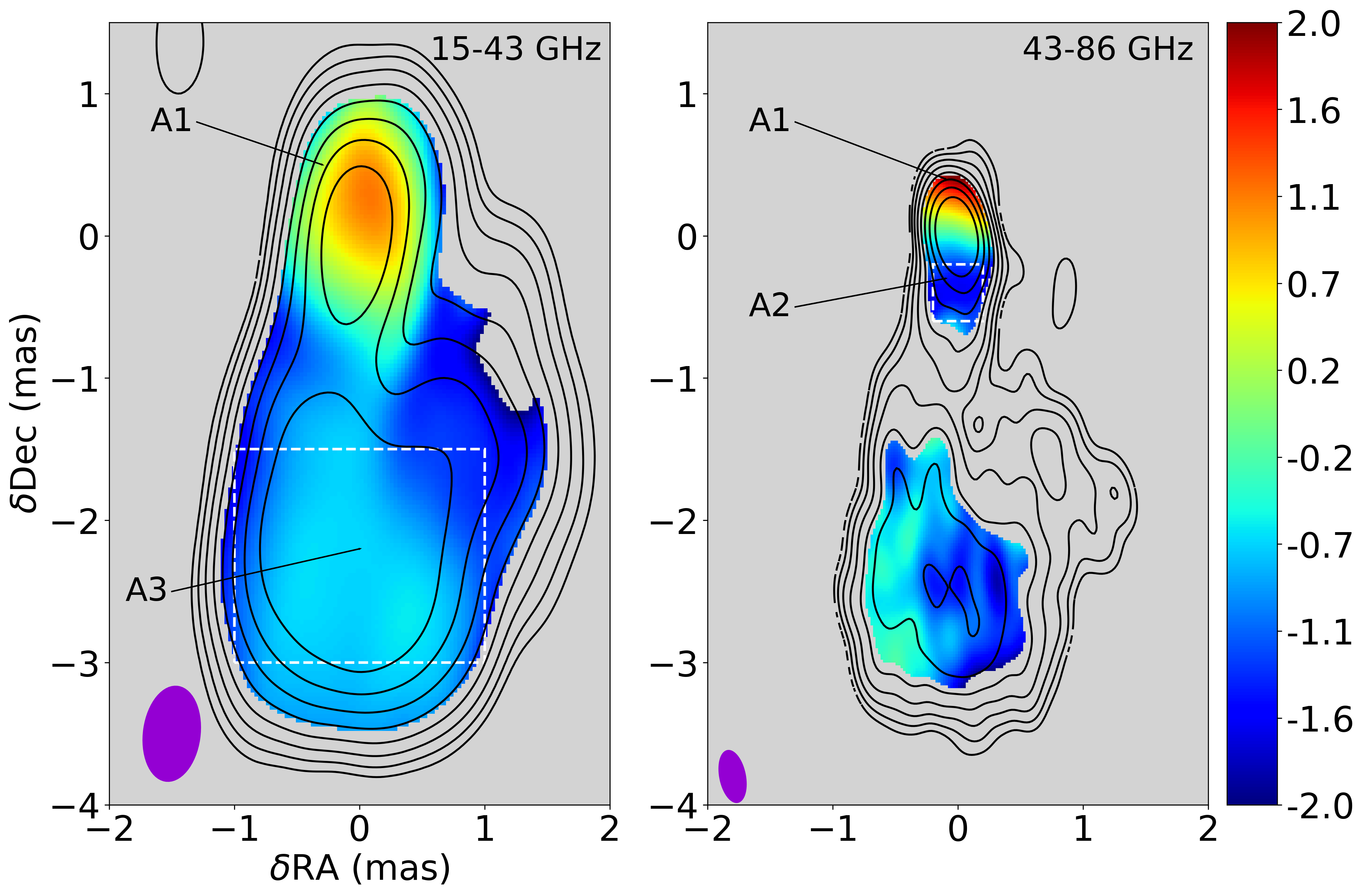}
   \caption{Core offset position fit and spectral index maps at 15-43\,GHz and 43-86\,GHz of \C. \emph{Left:} Core shift of \C\ at frequencies of 15, 43, and 86\,GHz. The red line shows a fit of Eq. \ref{eq:kr} to the core offset positions (for the numbers, see Table \ref{table:Absolute}). The shaded grey line denotes the distance from the 86\,GHz total intensity peak to the true location of the jet apex. The insets show 2D cross-correlation coefficients of the images at two different frequency pairs. Each axis is in units of mas. The contours start at 1\% and increase in steps of 10\% relative to the maximum cross-correlation coefficient (\cff\ and \cfe,\ respectively). The perpendicular dashed white lines intersect at the maximum of the colour map, which corresponds to the maximum value of $\rho(i, j)$. \emph{Right:} Spectral index maps after image alignment. The left panel shows the 15\,GHz image in contours and the 15-43\,GHz spectral index in colours. The contours start at 4.36\,mJy/beam and increase by a factor of two.  The maps are convolved with a beam of $0.45\times0.67$\,mas, oriented at a PA\,=\,-9.44$^{\circ}$. The right panel shows the 43-86\,GHz spectral index in colours and the contours of the 43\,GHz map, which start at 4.26\,mJy/beam and increase in steps of two.  The maps are convolved with a beam of $0.20\times0.37$\,mas, oriented at a PA\,=\, 14.1$^{\circ}$. We only display regions that have (i) a S/N of at least five for the total intensity contours and (ii) a spectral index in the range $-2$ to $2$, within $1$ rms error ($\pm0.14$ for the 15-43\,GHz pair and $\pm0.21$ for the 43-86\,GHz pair), gained from the uncertainty from a region the size of each beam, centred around the core shift location of each frequency (see Appendix \ref{App:Error} for further details). Boxes marked by dashed white lines correspond to the regions used for the 2D cross-correlation. A1, A2, and A3 are the three characteristic regions where the spectral indices were measured (see Table \ref{table:SpIMeas}).} 
    \label{fig:SpI}
    \end{figure*}

Following the analysis presented in \cite[][and references therein]{2014MNRAS.437.3396K}, we computed the core shift and magnetic field strength and estimated the topology of the magnetic field. For the analysis described below,
we convolved the maps with a circular beam with a size corresponding to the geometric mean of the major and minor axis of the lower-frequency beam (see Table \ref{table:Beams}). 
After the convolution of each image pair, we selected a region within the optically thin part of the jet for the alignment,  for which positions are frequency independent. 
For the 15-43\,GHz pair, we selected region A3, which is around the bright and well-defined
component C3 \citep{2014ApJ...785...53N}, located $\sim 2$\,mas south of the VLBI core (see Fig. \ref{fig:SpI}).
For the 43-86\,GHz pair, we aligned the maps using region A2, which is located
closer to the core (see Fig. \ref{fig:SpI}). A2 is bright enough and well defined, and it also shows a steep spectrum (see e.g. Fig. \ref{fig:SpI}), which justifies our choice.

The next step was to iteratively shift one image against the other in right ascension (RA) and in declination (Dec.). For each position shift $(i, j)$ the cross-correlation coefficient
was calculated as follows \citep{Lewis95}:
\begin{equation}
\resizebox{.45 \textwidth}{!}{$
\rho(i, j)= \frac{\sum_{x, y}\left[f^{\nu_1}(x, y)-\bar{f}^{\nu_1}_{i, j}\right]\left[f^{\nu_2}(x-i, y-j)-\bar{f}^{\nu_2}_{i, j}\right]}
{\left\{\sum_{x, y}\left[f^{\nu_1}(x, y)-\bar{f}^{\nu_1}_{i, j}\right]^{2} \sum_{x, y}\left[f^{\nu_2}(x-i, y-j)-\bar{f}^{\nu_2}_{i, j}\right]^{2}\right\}^{0.5}}
$}, \label{eq:cc}
\end{equation}
where \emph{x} and \emph{y} are the pixel indices in the different images, \emph{$f^{\nu_1}(x, y)$} is the flux density of the selected feature in the static image at $(x, y)$, \emph{$f^{\nu_2}(x-i, y-j)$} is the flux density of the selected feature in the shifted map (shifted by \emph{x-i} and \emph{y-j}), and \emph{$\bar{f}^{\nu_{1,2}}_{i, j}$} are the mean fluxes in the selected feature. The maximum $\rho(i, j)$ yields the best shift position. For further details
on the method, we refer to \cite{Lewis95}.

We aligned adjacent frequency maps pairwise to avoid artefacts due to the very different beam sizes at 15 and 86\,GHz.
In order to be as conservative as possible, we used the larger beam size at the longer wavelength (see Table \ref{table:Beams}) for the alignment of each frequency pair.
At 15-43\,GHz, we obtain position shifts of 
$\Delta $RA\,=\,\DRAfifteen\,$\mu$as and $\Delta$Dec.\,=\,\DDECfifteen\,$\mu$as.
At 43-86\,GHz, we obtain position shifts of  $\Delta$RA\,=\,\DRAfourtythree\,$\mu$as and $\Delta $Dec.\,=\,\DDECfourtythree\,$\mu$as. 
A description of our conservative error estimation is presented in Appendix \ref{App:Error}. Table \ref{table:Shifts} summarises the results.

We define the core of our images as the brightest and most compact component at the northernmost region of the jet.
As a preparatory step for
the 2D cross-correlation at 15 and 43\,GHz, we fitted
circular Gaussian components at 15\,GHz to the C3 and core region and determined their relative distance to be $2.24\,\pm\,0.02$\,mas. We then aligned the 15\,GHz and
43\,GHz maps by this shift and performed the cross-correlation. At  43\,GHz and
86\,GHz, the intensity peak is at the northernmost region of the jet and no additional shift is
necessary. 
In this procedure and at all frequencies, we fitted a circular Gaussian to the VLBI core to identify their position shifts from the phase centres, finding $<16\,\mu$as offsets at all frequencies (i.e. significantly smaller than the pixel scale and the image shifts). We thus ignore these offsets in the following analysis. 
The uncertainties of the image alignment for both frequency pairs is further discussed in Appendix \ref{App:Error}.

Figure \ref{fig:SpI} (left panel) displays the core shift values as a function of  frequency. We used the VLBI core at 86\,GHz as a reference, setting its position to zero. 
The two insets present the distribution of the cross-correlation coefficient for the 15-43\,GHz pair (left) and the 43-86\,GHz pair (right).

We then used this map alignment to calculate the spectral index distributions.
In Fig. \ref{fig:SpI} (right panel) we show the spectral index maps at 15-43\,GHz and at 43-86\,GHz. The northern side of the core region exhibits an inverted spectral index of $\alpha_{15-43}= 1.0 \pm 0.3$. The $\alpha_{15-43}$ gradually decreases southwards, with a
typical value of $\alpha_{15-43}= -0.5 \pm 0.4$ in the C3 region, consistent with past spectral index measurements at lower frequencies (e.g. \citealt{1982ApJ...256...83U, 1988Natur.334..131B, 1989ApJ...337..671M, 1994ApJ...430L..41V, 1994ApJ...430L..45W, 2006MNRAS.368.1500T, 2020ApJ...895...35W}).

For the 43-86\,GHz pair, we detect a  prominent spectral index gradient between regions
A1 and A2 (see Fig. \ref{fig:SpI}), from $\alpha_{43-86}= 2.0 \pm 0.5$ in the north-west to 
$\alpha_{43-86} =  -1.0 \pm 0.6$ in the south-east. 
Details on the spectral index error estimation are given in Appendix \ref{App:Error}.
Based on this, we conclude that the overall trend of the spectral index gradients from inverted in the northern region (A1) to steeper in the two southern regions (A2 and A3) are significant and real. We also note that the apparent difference between the spectral index in the two southern regions (A2: $\alpha \sim -1.0$; A3: $\alpha \sim -0.5$) may not be significant, due to residual calibration uncertainties. However, due to the
nature of C3 (moving shock) and its interactions with the ambient jet, the flatter spectrum in region A3 is not unexpected (e.g. \citealt{2014ApJ...785...53N}).

\begin{figure}
   \centering
   \includegraphics[scale=0.3]{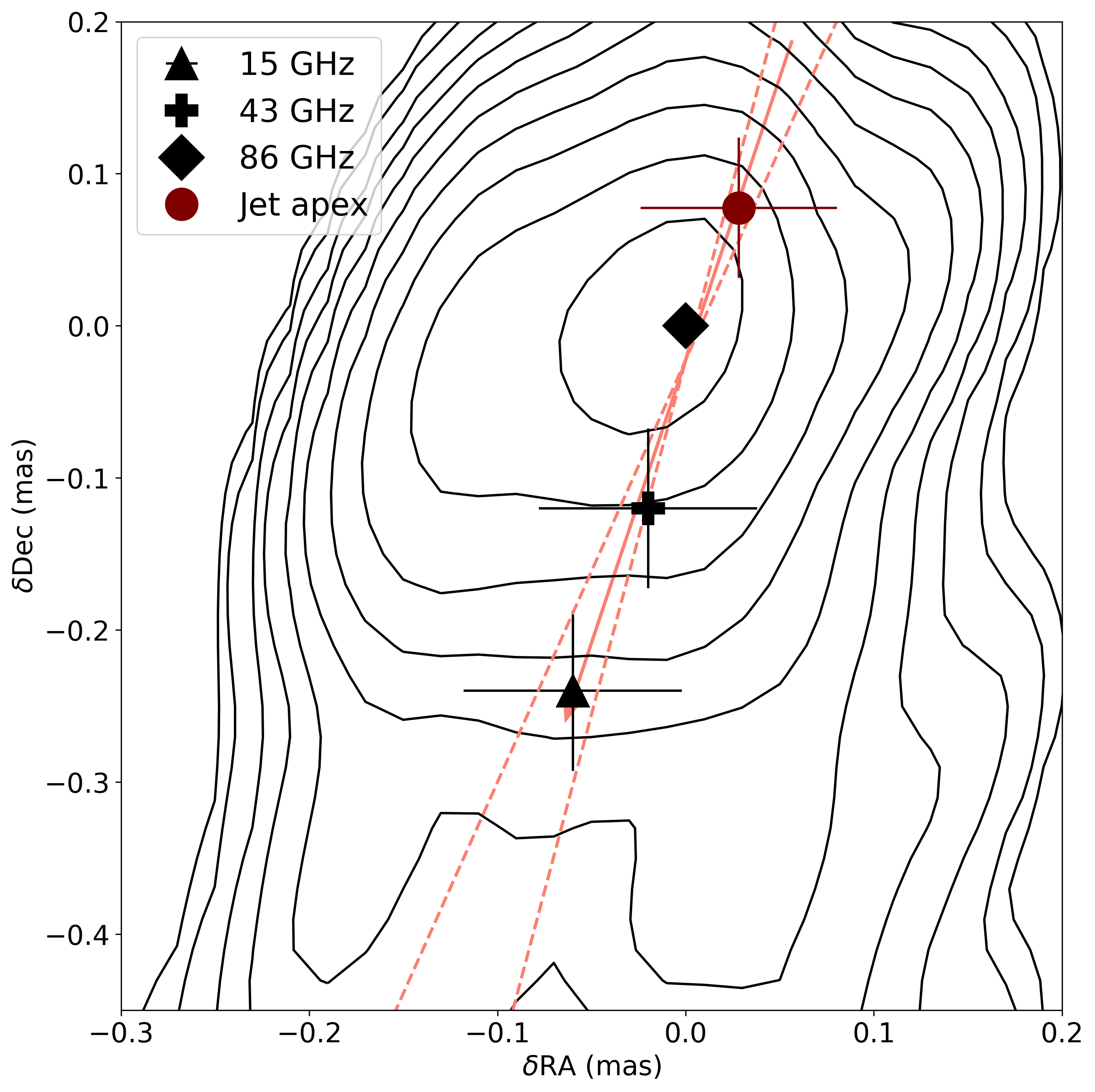}
      \caption{ Core locations at 15, 43, and 86\,GHz (black) and extrapolated jet apex location (dark red), plotted over the 86\,GHz total intensity map. The contours start at 0.1\% of the image peak (1.82\,Jy/beam) and then increase in steps of two. The solid light red line is obtained from a 2D line fit to the core positions at different frequencies and the jet apex location with respect to the reference point (i.e. 86\,GHz core).\ The broken light red lines show the 99\% confidence interval of the fit.}
         \label{fig:Peak}
   \end{figure}

If the size of the emission region increases linearly with the distance, $r$, from the core, $w\propto r$, the magnetic field strength and particle density decrease with distance as $B\propto r^{-b}$ and $N\propto r^{-n}$, respectively. The distance between the jet apex and the apparent VLBI core, $\Delta r_{\rm core}$, relates to the frequency with a power law of the form:
\begin{equation}
    \Delta r_{\rm core}=r_0\left(\left(\frac{\nu}{\rm 86\,GHz}\right)^{-1/k_r}-1\right)\ \text{[$\mu$as]} \label{eq:kr}
,\end{equation}
where $k_r=[2n+b(3-2\alpha_{\rm thin})-2]/(5-2\alpha_{\rm thin})$ \citep{1998A&A...330...79L}, $n$ and $b$ are the particle number density and magnetic field strength power-law indices, respectively, and $\alpha_{\rm thin}$ is the optically thin spectral index. 
According to the definition of Eq. \ref{eq:kr}, and using the 86\,GHz core as the reference point, we can determine $|r_0|$, which is the distance to the jet apex.
To obtain the absolute distances of the 15\,GHz and the 43\,GHz cores from the 86\,GHz core, we added the $\Delta$RA and $\Delta$Dec. shifts listed in Table \ref{table:Shifts} in quadrature. Table \ref{table:Absolute} summarises the resulting positions.
The limited frequency range and scarcity of the data points only allow a fit for one free parameter, namely $r_0$. We therefore assumed a physically motivated value of $k_r=1$, as also observed in other VLBI jets  \citep[e.g.][]{1998A&A...330...79L, 2011Natur.477..185H, 2012A&A...545A.113P, 2013A&A...557A.105F}. 
We solved Eq. \ref{eq:kr}  via the least squares fit method for $r_0$. 
We used the inverse variance of the data for their weighting.
From the fit we obtain $r_0=\propconstalt$ $\mu$as, 
which is marked as a grey line in Fig. \ref{fig:SpI} (left)
\footnote{We note the good agreement with the work of \cite{Oh21}, who find an offset of $54-215$\,$\mu$as from long-term, time-averaged 86\,GHz GMVA imaging.}.
This corresponds to a projected distance of \Projpc\,pc, or \ProjRs\,R$_s$ (with the Schwarzschild radius, R$_s$). Adopting a viewing angle in  the range of 20$^\circ-65^\circ$ \citep{2017MNRAS.465L..94F, 2009ApJ...699...31A}, we determine the de-projected distance to be \Dpc\,pc, or \DRs\,R$_s$. 
For the position angle (PA) of the jet apex, we obtain \PA ~relative to the 86\,GHz core position in the image plane.
In Fig. \ref{fig:Peak} we over-plot the core positions at 15, 43, and 86\,GHz, as well as the estimated position of the jet apex, on the intensity contours of the 86\,GHz map. The estimated position of the jet apex, marked as a filled circle, is located north-west of the VLBI core at 86\,GHz.

\section{Discussion}\label{sec:Disc}

\subsection{Implications of the jet apex location}

Over the years, a variety of interpretations have been proposed to explain the highly complex structure in \C. Based on 22\,GHz space-VLBI imaging with $20\,\mu$as resolution, 
\cite{2018NatAs...2..472G} found an east-west, broadly elongated structure at
the northern end of the jet. The authors suggest that this may correspond to a 
broad jet apex of $\sim120\,$R$_{s}$ width, with some diffuse emission farther to the north, marking the onset of the counter-jet. 
 Our analysis indicates a different scenario, in which the jet apex is located more upstream, at $\sim$\,\DRs\,R$_s$ 
north-west of the VLBI core at 86\,GHz, as illustrated in Fig. \ref{fig:Peak}.
The measured orientation along PA\,=\,\PA\ favours this scenario because it is in line with the direction of the northern counter-jet PA = $-25^{\circ}$, as seen on the larger mas scales \citep{1994ApJ...430L..41V}.

We point out that on sub-mas scales the northern emission from the counter-jet is not yet unambiguously detected. It is unclear if the emission seen north of the brightest jet components at 43\,GHz and 86\,GHz belongs to the jet or to the counter-jet.
\cite{2017MNRAS.465L..94F} report the detection of a counter-jet at 15 and 43\,GHz in the
northern 1-2\,mas region. The apparent emission gap between this mas-scale region and the sub-mas-scale region close to the VLBI core (as seen at higher frequencies) 
may be explained by free-free absorption from optically thick and clumpy gas, for example a torus or accretion flow 
(e.g. \citealt{2008A&A...483..793S, 2017MNRAS.465L..94F, 2019A&A...622A.196K}). The highly inverted spectrum ($\alpha_{43-86}\sim2$ for the core region; see Fig. \ref{fig:SpI}, right panel) could support such a  scenario. However, highly inverted spectra (with spectral indices
reaching up to $\alpha=+2.5$) are also possible for homogeneous self-absorbed synchrotron emitting components. Depending on homogeneity, the VLBI cores in many AGN jets more typically show inverted spectra with lower indices in the range of 0 to 1.5, though occasional higher values cannot be excluded. Dedicated spectral and Faraday rotation measure-sensitive polarimetric millimetre-VLBI observations in the future
could help to clarify the situation.

\subsection{Magnetic field strength and topology}

Following \cite{1998A&A...330...79L} and Eqs. 38 and 39 in  \cite{2013A&A...557A.105F}, we determined the magnetic field strength in the jet.
We assumed a power-law index $k_r=\kralt$. Details of the procedure and parameter ranges are given in Appendix \ref{App:Bfield}. Table \ref{table:Fits} summarises the relevant parameters. 
For the location of the jet apex with respect to the 86\,GHz core ($r_0=\coreshift\,\mu$as), we computed the magnetic field to be in the range of $B=\Bzero$\,G at the 86\,GHz core; this is lower than the magnetic field strength estimate of $21\pm14$\,G obtained by \cite{2019A&A...622A.196K}, where the synchrotron self-absorption formula from \cite{1983ApJ...264..296M} was used with several  assumptions. 
We followed the literature (\citealt{2013A&A...557A.105F, 2014MNRAS.437.3396K, 2017MNRAS.468.4478L} and references therein) to determine the magnetic field topology and thus assumed a power-law index of $n=2$ for the radial dependence of the particle number density. 
From this, we derive $b=\bindexalt$ for the power-law index of the magnetic field, which is commonly interpreted as evidence for a toroidal magnetic field configuration. 

We note that the above analysis of the magnetic field strength (see also Appendix \ref{App:Bfield}) largely relies on the \cite{1979ApJ...232...34B} jet model, where the particle number density and the magnetic field strength gradients depend on the shape of the jet (see also \citealt{1998A&A...330...79L}).
Thus, we can further examine our assumption by adopting a more realistic boundary shape of the jet and checking the dependence of $b$ on $n$.
That is, the radius (or transverse width) of a compact jet, $w$, depends on the core distance, $r$. Previous such studies of \C\ \citep{2014ApJ...785...53N, 2018NatAs...2..472G} revealed that, on average,
$w\propto r^{0.21}$. Thus, the jet cross-sectional area $A(r)\propto w^{2} \propto r^{0.42}$. The total number of particles, $N_{\rm tot}$, passing through each cross-section is assumed to be conserved. Therefore, assuming a slab of width $dr$, the total number of particle $N_{\rm tot} \propto N(r)A(r)dr$\footnote{We further  assume a constant jet speed over the short distance scales discussed in this paper.} needs to be constant, where $N(r)\propto r^{-n}$. This leads to the number density $N(r) \propto A(r)^{-1}$ or $N(r)\propto r^{-0.42}$, and thus we now have $n=0.42$. 
We then obtain $b\bindexthreealt$, which comes closer to a poloidal magnetic field configuration ($b=2$). However, the underlying uncertainty in $b\bindexthreealt$ could be large due to the $k_1=1$ assumption. Future millimetre-VLBI imaging of the electric vector polarisation angle (EVPA) distributions in the core region would independently and directly measure the $b$ parameter.

We also note that the jet apex and the location of the central engine (the SMBH) may not coincide spatially. That is, the jet apex, which is the upstream end of a luminous flow of relativistic plasma, can only be physically associated with the central engine when the jet base emits synchrotron radiation. 
Therefore, the conversion of the magnetic or Poynting-flux energy into particle energy
at the jet base (e.g. `magnetoluminescence'; \citealt{blandford17}) must be efficient enough. If this is not the case, the `physical' origin of the jet (the location of the SMBH) may even be located beyond the position of the jet apex derived above.

\cite{1982MNRAS.199..883B} and  \cite{1977MNRAS.179..433B} presented two different jet launching scenarios. 
In the former the jet is powered by magnetic field lines anchored in the accretion disk, while in the latter the magnetic field lines are directly connected to the ergosphere of the spinning black hole.
Depending on the chosen value of $n$, we conclude that the magnetic field configuration
is either a toroidal or is of a mixed toroidal-poloidal nature.
Such a scenario is also consistent with observational findings, which indicate the presence of toroidal fields in AGN jets from scales of parsecs to kiloparsecs \citep{2014A&A...566A..26M, 2017Galax...5...61K}.
Under the assumption of the validity of the \cite{1979ApJ...232...34B} jet model,
our data suggest that, at a distance of $\sim$\,\DRs\,R$_s$ from the jet apex,
the magnetic field configuration is most likely mixed. We note that the initial magnetic field configuration of the BP model is expected to be toroidal, whereas the BZ model predicts a poloidal geometry \citep{2020ARA&A..58..407D}. Therefore, the observed mixed configuration either points to a stratified combination of both the BP and BZ models or to a jet launching process in which the initial field configuration is altered by some internal physical jet processes acting farther downstream, such as developing shocks and/or instabilities.

To date, the magnetic field strength of only a few other nearby AGN has been studied on scales
of a few hundred R$_{s}$.
 \cite{2016A&A...593A..47B} determined the magnetic field strength of NGC\,1052 to be $\geq 100$\,G at $\sim 4$\,R$_s$. \cite{2018A&A...616A.188K} computed the magnetic field of M\,87 to be in the range of $\sim60-210$\,G on $\sim10$\,R$_s$ scales. 

Adopting $b=1$ and extrapolating to 10\,R$_s$, the magnetic field of \C\ is of the order of $70-600$\,G and thus compares well to those in NGC\,1052 and M\,87.
Furthermore, we can extrapolate the $B_0$ of \C\ to a distance of 1\,pc from the jet apex, $B_{\rm 1pc}$, in order to compare this value to literature results of additional AGN jets from a time when such high spatial resolution imaging was not yet possible. Using $b=1$, we obtain $B_{\rm 1pc}$ to be \Bone\,mG. 
\cite{2012A&A...545A.113P} obtain $B_{\rm 1pc} \sim  400-900$\,mG for a total of $\sim100$ jets in quasars and BL Lac objects based on their core shifts. 
The B field found in \C\ is a factor of four to six lower, which\ may indicate intrinsic differences between radio galaxies and the more luminous quasars and BL Lac objects.

It is interesting to compare our result to the magnetic field strength expected from the total jet power. Using the following equation (see Eq. 8.35 in \citealt{2013LNP...873.....G} and the listed assumptions),

\begin{equation}
    B\leq\sqrt{\frac{8\pi P}{A\Gamma^2\beta c}} \label{eq:Power}
,\end{equation}
where $P$ is the total jet power, $\beta$ is the intrinsic speed of the jet, $\Gamma$ is the associated bulk Lorentz factor, and $A$ is the total cross-section of the jet, we can compute an estimate for the magnetic field strength from the total jet power for \C. We note that the equality holds only if the total energy budget of the jet is solely dominated by the magnetic field. \cite{2009ApJ...699...31A} and \cite{2018A&A...617A..91M} report a jet power as high as $\sim10^{44}-10^{45}$\,erg/s, stemming from the observed extreme teraelectronvolt $\gamma$-ray variability. 
Thus, for an emitting region the size of the jet cross-sectional area ($\sim50\,\mu$as), a magnetic field strength of $\lesssim$\,\Bfromjetpower\,G would be possible without exceeding the jet power. Our finding for the magnetic field strength does not exceed this upper limit.  The moderately large magnetic field strength from the core shift, in comparison to the upper limit estimated from the jet power analysis, 
therefore also
supports a scenario in which  the magnetic field at the jet base is prominent, which, in turn, is in support of magnetic jet launching (e.g. \citealt{2003PASJ...55L..69N, 2011MNRAS.418L..79T}).

\section{Conclusions} \label{sec:Conclusions}

In this letter we have studied the spectral index and core shift of \C. Our major findings and conclusions can be summarised as follows.
\begin{enumerate}
    \item We performed a 2D cross-correlation of the 15-43\,GHz and 43-86\,GHz image pairs, using quasi-simultaneous VLBI observations from May 2015. The analysis of the core shift reveals that the jet apex is located north-west of the VLBI core at 86\,GHz, displaced by \coreshift\,$\mu$as, with the distance from the core as a function of the frequency, following Eq. \ref{eq:kr}.
    \item With the detected core shift and the possible location of the jet apex north of the 86\,GHz VLBI core,  the east-west oriented VLBI core structure (which is also seen in the {22\,GHz \it RadioAstron} map) appears less likely to be the physical origin of the jet. We further note that a location of the true jet apex north of this east-west oriented feature and north of the 86\,GHz $\tau\sim1$ surface would also lead to a smaller initial jet-opening angle (as opposed to the $130^\circ$ angle that has been suggested by \citealt{2018NatAs...2..472G}).
    \item The new spectral index images at 15-43\,GHz and, especially, 43-86\,GHz reveal the presence of a strong spectral index gradient in the northwest-southeast direction,
    with an inverted spectrum of the millimetre-VLBI core ($\alpha_{43-86} \sim +2$). With a synchrotron turnover frequency of $\nu_m \geq 86$\, GHz, \C\ will be a suitable
    target for VLBI studies at higher frequencies (e.g. with the Event Horizon Telescope, EHT; \citealt{2019ApJ...875L...2E}).

    \item At a de-projected distance of \DRs\,R$_s$ (\coreshift\,$\mu$as) from the jet apex, 
    the magnetic field topology is not purely poloidal; a mixed poloidal-toroidal
    configuration is suggested. This points towards a stratified combination of the BP and BZ models (acting in parallel) or towards an alteration to the initial magnetic field configuration due to some internal physical jet processes acting farther downstream (e.g. developing shocks and/or instabilities).
    \item We measure the magnetic field to be in the range $B_0=$\,\Bzero\,G at the jet apex. This value is lower compared to the maximum possible magnetic field strength derived from the total jet power, which is $\lesssim$\,\Bfromjetpower\,G.  
    The magnetic field also compares well with the magnetic field measured in some other nearby AGN, such as M\,87 and NGC\,1052, and suggests magnetic jet launching.
\end{enumerate}

\noindent
Overall, our study suggests that the complex nature of \C\ can be partially explained by the location of the jet apex being upstream from the VLBI core at 86\,GHz. Questions about the nature of the east-west oriented elongated VLBI core, including whether it is a stationary or oblique shock or part of a curved filament in a wider jet channel, still remain open.  A broader frequency coverage from quasi-simultaneous observations may be necessary to achieve an improved estimate of the power-law index ($k_r$) of the core shift.
We plan to further investigate, and constrain our results, by employing millimetre-VLBI monitoring observations with the highest possible resolution (EHT, GMVA, Global-EVN) in the near future.

\begin{acknowledgements}
      We thank T. Savolainen for providing software to calculate two dimensional cross-correlations.  We also thank N. R. MacDonald for the proofreading and fruitful discussions which helped improve this manuscript. We thank the anonymous referee for the valuable comments, which improved this manuscript.
      G. F. Paraschos is supported for this research by the International Max-Planck Research School (IMPRS) for Astronomy and Astrophysics at the University of Bonn and Cologne. 
      This research has made use of data obtained with the Global Millimeter VLBI Array (GMVA), which consists of telescopes operated by the MPIfR, IRAM, Onsala, Mets\"ahovi, Yebes, the Korean VLBI Network, the Green Bank Observatory and the Long Baseline Observatory. The VLBA is an instrument of the Long Baseline Observatory, which is a facility of the National Science Foundation operated by Associated Universities, Inc. The data were correlated at the correlator of the MPIfR in Bonn, Germany. This work makes use of the Swinburne University of Technology software correlator, developed as part of the Australian Major National Research Facilities Programme and operated under licence. This study makes use of 43\,GHz VLBA data from the VLBA-BU Blazar Monitoring Program (VLBA-BU-BLAZAR; \url{http://www.bu.edu/blazars/VLBAproject.html}), funded by NASA through the Fermi Guest Investigator Program. This research has made use of data from the MOJAVE database that is maintained by the MOJAVE team \citep{2009AJ....137.3718L}. This research has made use of the NASA/IPAC Extragalactic Database (NED), which is operated by the Jet Propulsion Laboratory, California Institute of Technology, under contract with the National Aeronautics and Space Administration. This research has also made use of NASA's Astrophysics Data System Bibliographic Services. Finally, this research made use of the following python packages: {\it numpy} \citep{harris2020array}, {\it scipy} \citep{2020SciPy-NMeth}, {\it matplotlib} \citep{Hunter:2007}, {\it astropy} \citep{2013A&A...558A..33A, 2018AJ....156..123A} and {\it Uncertainties: a Python package for calculations with uncertainties}.
\end{acknowledgements}

\bibliographystyle{aa} 
\bibliography{aanda} 

\begin{appendix} 

\section{Error estimation}\label{App:Error}
The total error budget of our cross-correlation analysis depends on the cross-correlation image alignment error ($\sigma_{\rm ccf}$), the image parameters, the correlation region, and the rms error ($\sigma_{\rm core}$) of the core shift \citep{2017MNRAS.468.4478L}. 
The image size, convolving beam size, and pixel scale should be varied with frequency. Before we performed a cross-correlation of two images, these three parameters needed to be set to a common value. 
To consolidate the image parameter differences, we chose the beam size and map size of the lower-frequency map and the pixel scale of the higher-frequency map.
We found minor offsets, which were taken as the uncertainties for the image alignment, as described in the next paragraph.
The error of the core position
($\sigma_{\rm core}$) depends primarily on the maximum resolution, $d_{\rm lim}$, that can be achieved for a given map, which is given by Eq. \ref{eq:Lob05} and taken from \cite{2005astro.ph..3225L}: 

\begin{equation}
d_{\lim }=\left[\frac{16}{\pi} \Theta^2 \ln 2 \ln \left(\frac{S/N}{S/N-1}\right)\right]^{1 / 2}
[\text{mas}]. \label{eq:Lob05} 
\end{equation} 
Here, $\Theta$ is the radius of the circular beam and S/N is the signal-to-noise ratio. For the 15-43\,GHz pair, we calculated the S/N from the ratio of the image peak and rms error  to be $\sim500$ by averaging the S/Ns from the 15 and 43\,GHz images. We followed the same procedure for the 43-86\,GHz pair and determined the S/N to be $\sim375$. We followed \cite{2012A&A...537A..70S} and used Eq. (7) therein:

\begin{equation}
\sigma_{\rm core}=\frac{\mathfrak{D}}{\rm S/N}, \label{eq:Schinzel12} 
\end{equation} 
where $\mathfrak{D}$ is either equal to $d_{\rm lim}$ or to the central component size, $d,$ if $d_{\rm lim}<d$. 
In our case $d_{\rm lim}>d$, so the first case applied. 
The final values for the error budget due to the uncertainty of the core position, $\sigma_{\rm core}$, are $\sigma_{\rm core}^{15-43}=\imageff$\,$\mu$as and $\sigma_{\rm core}^{43-86}=\imagefe$\,$\mu$as.

Subsequently, we adhered to the following procedure to calculate the cross-correlation image alignment error, $\sigma_{\rm ccf}$. First, two of the authors did independent imaging of the three available frequency maps to produce different sets of images. We then aligned them using the similar, though not identical, optically thin regions and observed the difference in the alignment. Then, we used half of the shift differences as the image alignment uncertainty. The resulting uncertainties are $\sigma_{\rm ccf}^{\rm RA} = \RAuncertaintyff$\,$\mu$as and $\sigma_{\rm ccf}^{\rm Dec.} = \DECuncertaintyff$\,$\mu$as in the RA and Dec. directions for the 15-43\,GHz pair and $\sigma_{\rm ccf}^{\rm RA} = \RAuncertaintyfe$\,$\mu$as and $\sigma_{\rm ccf}^{\rm Dec.} = \DECuncertaintyfe$\,$\mu$as for the 43-86\,GHz pair, respectively.
For the final values for each image, we added $\sigma_{\rm core}$ to $\sigma_{\rm ccf}$ in quadrature,  thus $\sigma_{\rm tot} = \sqrt{ \sigma_{\rm core}^2 + \sigma_{\rm ccf}^2 }$, and used $\sigma_{\rm tot}$ to calculate the errors shown in Table \ref{table:Absolute}.

We also estimated an error budget for our spectral index map by creating spectral index error maps. We used a combination of the image noise uncertainty and systematic amplitude calibration error to derive the error budget, as presented in Eq. (3) of \cite{2014JKAS...47..195K}. 
The image alignment uncertainty, as described in the previous paragraph, yields  uncertainties of $\sim0.3$ for the 15-43\,GHz pair and $\sim0.5$ for the 43-86\,GHz pair. For the 15-43\,GHz pair, we find a  systematic amplitude calibration error of 0.30 and for the 43-86\,GHz pair 0.35; both these values refer to the entire region of emission.  Adding these in quadrature produces total error estimates of $\Delta\alpha^{\rm entire}_{15-43}=0.4$ and $\Delta\alpha^{\rm entire}_{43-86}=0.6$. We also focused on the region around the core, this time examining only an area of the beam size for each pair, centred on the location of the core shift, where we found the spectral index gradient. The 15-43\,GHz pair yields a systematic amplitude calibration error of 0.14, and the 43-86\,GHz pair 0.21. 
Adding again the error from the alignment in quadrature to these values yields total error estimates of $\Delta\alpha^{\rm core}_{15-43}=0.3$ and $\Delta\alpha^{\rm core}_{43-86}=0.5$.
For the three regions, the spectral indices and their errors are summarised in Table \ref{table:SpIMeas}.

\begin{table}
\caption{Summary of the image parameters.}             
\label{table:Beams}      
\centering                          
\begin{tabular}{c c c}        
\hline\hline                 
Image & Beam [mas, deg] & Pixel scale [$\mu$as/pixel] \\    
\hline                        
   Fig. \ref{fig:SpI}, left & $0.45\times0.67$ $(-9.44)$ & 30 \\      
   Fig. \ref{fig:SpI}, right & $0.20\times0.37$ $(14.1)$ & 20 \\
   Fig. \ref{fig:SpI_zoomed}, left & $0.55\times0.55$ $(0)$ & 30 \\
   Fig. \ref{fig:SpI_zoomed}, right & $0.27\times0.27$ $(0)$ & 20 \\
\hline                                   
\end{tabular}
\end{table}

\begin{table}
\caption{Relative core shifts for the low- and high-frequency alignment pairs.}           
\label{table:Shifts}      
\centering                          
\begin{tabular}{c c c}        
\hline\hline                 
$\nu_{pair}$ [GHz] & $\Delta$RA [$\mu$as] & $\Delta$Dec [$\mu$as] \\    
\hline                        
   15\,-\,43 & \DRAfifteen & \DDECfifteen \\      
   43\,-\,86 & \DRAfourtythree & \DDECfourtythree \\
\hline                                   
\end{tabular}
\end{table} 

\begin{table}
\caption{Absolute distances from the extrapolated jet apex.}             
\label{table:Absolute}      
\centering                          
\begin{tabular}{c c}        
\hline\hline                 
$\nu$ [GHz] & $\Delta r$ [$\mu$as] \\    
\hline                        
   15 & \Absff \\      
   43 & \Absef \\
   86 & \propconst \\
\hline                                   
\end{tabular}
\end{table}


\begin{table}
\caption{Spectral index measurements per region.}           
\label{table:SpIMeas}      
\centering                          
\begin{tabular}{c c c}        
\hline\hline                 
Region & $\alpha_{\rm 15-43}$ & $\alpha_{\rm 43-86}$ \\    
\hline                        
   A1 & $1.0\pm0.3$ & $2.0\pm0.5$ \\      
   A2 & -- & $-1.0\pm0.6$ \\
   A3 & $-0.5\pm0.4$ & -- \\
\hline                                   
\end{tabular}
\end{table}


\begin{table}
\caption{List of the fit parameters that appear in the text.}             
\label{table:Fits}      
\centering                          
\begin{tabular}{c c}        
\hline\hline                 
Parameter & Value \\    
\hline                        
   $k_r$ & \kr \\      
   $r_0$ & \propconst\ [$\mu$as GHz] \\
   $b$ & \bindex\\
   $\Delta r_{\mathrm{core}}$ & \coreshift\ [$\mu$as]\\
   $\alpha_{\rm thin}$ & $-0.77$\\
   $\gamma_{\rm max}$ & $10^{3}-10^{5}$\\
   PA & \PA\\
   $\beta_{\rm app}$ & \bapp \\
   $\theta$ & 20${}^{\circ}-65{}^{\circ}$\\
   $\phi$ & \OpenAngle \\
   $\Omega^{\nu}_{r}$ & \Omegarnurange\ [pc\,GHz] \\
   $B_0$ &\Bzero\ [G] \\
\hline                                   
\end{tabular}
\end{table}

   \begin{figure}
     \centering
   \includegraphics[scale=0.25]{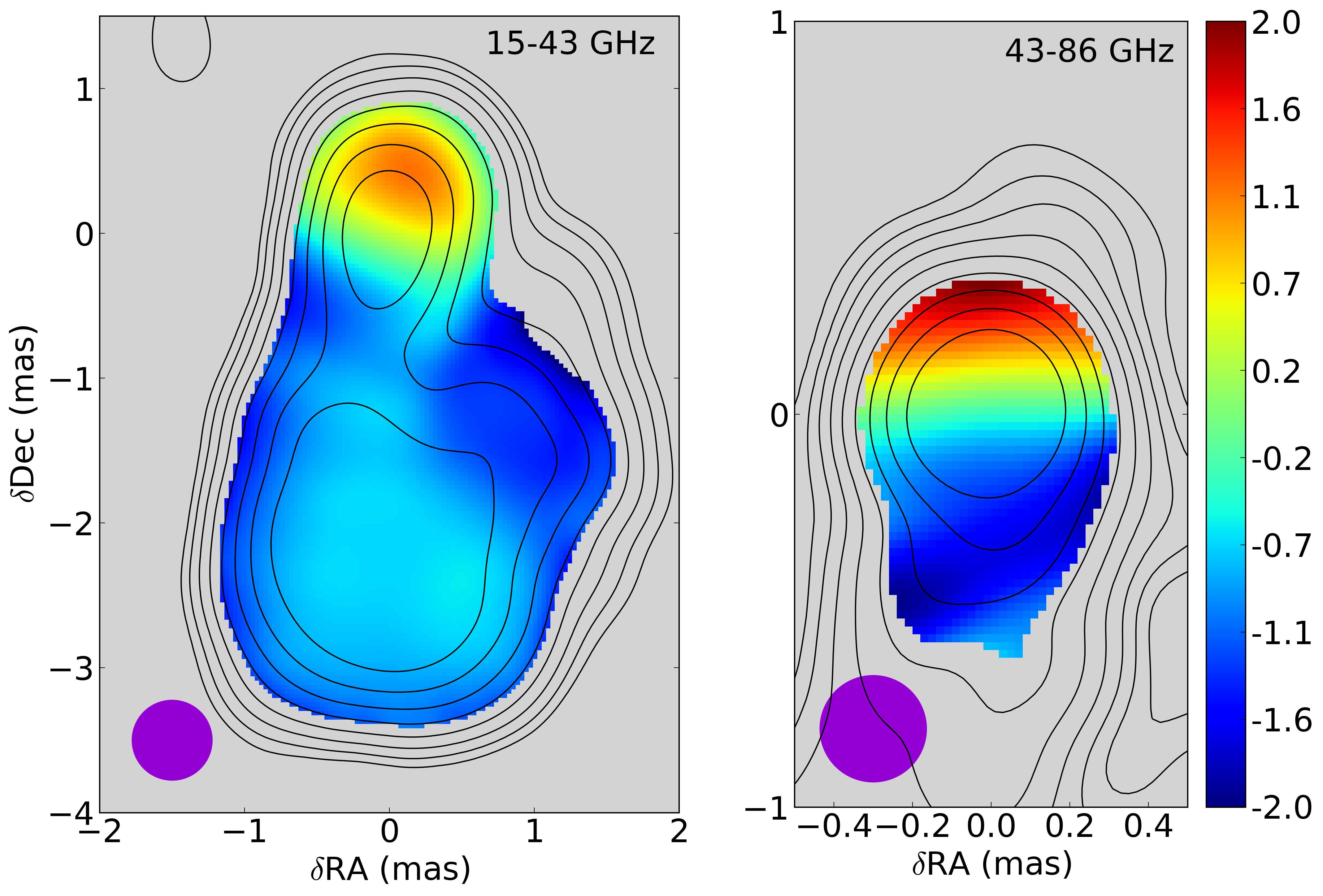}
   \caption{Spectral index maps after image alignment, convolved with a beam of a radius equal to the geometric mean of the lower-frequency map beam (see Table \ref{table:Beams}). The left panel shows the 15\,GHz image in contours and the 15-43\,GHz spectral index in colours. The contours start at 4.41\,mJy/beam and increase with a factor of two.  The spectral index map is convolved with a circular beam of $0.55$\,mas. For the right panel, we zoomed in on the nuclear region of the 43-86\,GHz spectral index image, shown in colour to better illustrate the spectral index gradient. The contours start at 4.38\,mJy/beam, corresponding to the 43\,GHz image, and increase in steps of two.  The spectral index map is convolved with a circular beam of $0.27$\,mas. We apply the same cutoffs as in Fig. \ref{fig:SpI}.}
    \label{fig:SpI_zoomed}
   \end{figure}

\section{Magnetic field strength estimation} \label{App:Bfield}

Following \cite{1998A&A...330...79L}, the core shift (in mas) between two frequencies, $\nu_1$ and $\nu_2$ (in GHz), with $\nu_1<\nu_2$, can be expressed in terms of the parameter $\Omega^{\nu}_{r}$:

\begin{equation}
\Omega^{\nu}_{r}=4.85 \times 10^{-9} \frac{\Delta r_{\mathrm{core}} D_{\mathrm{L}}}{(1+z)^{2}}\left(\frac{\nu_{1}^{1 / k_r} \nu_{2}^{1 / k_r}}{\nu_{2}^{1 / k_r}-\nu_{1}^{1 / k_r}}\right) [\text{pc GHz}{}^{1/k_r}],\label{eq:Omega}
\end{equation}
where $\Delta r_{\mathrm{core}}$ is the core shift in mas, $D_{\rm L}$ is the luminosity distance in pc, and $z$ is the redshift. We find that $\Omega^{15-43}_{r}=$\,\Omegarnuone\,[pc GHz] and $\Omega^{43-86}_{r}=$\,\Omegarnutwo\,[pc GHz]. 
For our subsequent analysis, we used a weighted mean of the two, yielding  $\Omega_{r}=\Omegarnualt$\,[pc GHz].
Equation (39) in \cite{2013A&A...557A.105F}, in turn, relates the magnetic field strength at a specified distance (in pc) from the jet base:

\begin{equation}
\begin{split}
B_{0} \approx \frac{2 \pi m_{\mathrm{e}}^{2} c^{4}}{e^{3}} &\left[\frac{e^{2}}{m_{\mathrm{e}} c^{3}}\left(\frac{\Omega^{\nu}_{r}}{r_{0} \sin \vartheta}\right)^{k_{r}}\right]^{\frac{5-2 \alpha_{0}}{7-2 \alpha_{0}}}\left[\pi C\left(\alpha_{0}\right) \frac{r_{0} m_{\mathrm{e}} c^{2}}{e^{2}}\frac{-2 \alpha_{0}}{\gamma_{\min }^{2 \alpha_{0}+1}}\right.\\
&\left.\times\frac{\varphi}{\sin \vartheta} 
K\left(\gamma, \alpha_0\right)
\left(\frac{\delta}{1+z}\right)^{\frac{3}{2}-\alpha_{0}}\right]^{\frac{-2}{7-2 \alpha_{0}}} [\text{G}],\label{eq:Beta}
\end{split}
\end{equation}
with
\begin{equation}
K\left(\gamma, \alpha_0\right) = \frac{2 \alpha_{0}+1}{2 \alpha_{0}} \frac{\left[\left(\gamma_{\max } / \gamma_{\min }\right)^{2 \alpha}-1\right]}{\left[\left(\gamma_{\max } / \gamma_{\min }\right)^{2 \alpha+1}-1\right]}
\end{equation}
and $C\left(\alpha_{0}\right)$ being tabulated in \cite{2005ApJ...619...73H} and $r_0$ a fixed distance so that $B=B_0(r_0/r)^b$. We set $r_{0}$ to be \coreshift\,$\mu$as, which is the distance from the 86\,GHz VLBI core to the jet apex. Furthermore, $\gamma_{\rm min}$ and $\gamma_{\rm max}$ are the minimum and maximum Lorentz factors for emitting electrons, $\theta$ is again the jet viewing angle, $\delta\equiv[\Gamma(1-\beta\cos(\theta))]^{-1}$ is the Doppler factor, $\Gamma= (1-\beta^2)^{-\frac{1}{2}}$ is the bulk Lorentz factor, and $\phi$ is the jet half opening angle. For $\gamma_{\rm max}$
we used the range $\gamma_{\rm max }=10^{3}-10^{5}$, reported by \cite{2009ApJ...699...31A}, and for $\gamma_{\rm min}$ we used a low value of 1. 
The jet speed and apparent jet speed are connected through Eq. (3.37) in \cite{2013LNP...873.....G}. For the jet half opening angle we used the formula $\phi=\arctan\left( \sin\left(\theta\right)\tan\left(\phi_{\rm app}/2\right) \right)$ from \cite{2017MNRAS.468.4992P}, having adopted the apparent opening angle range of $\phi_{\rm app}=\OpenAnglealt$\footnote{We set as an upper limit the angle created by the upper and lower limit of the PA of the core shift, as displayed in Fig. \ref{fig:Peak}. 
For the lower limit we used the full width at half maximum (FWHM) value of the bright component at a distance of $\sim2$\,mas from the VLBI core, as seen, for example, in Fig. \ref{fig:SpI}.}. 
For the jet speed near the VLBI core, a typical value range of $\beta_{\rm app}=\bapp$ for \C\ 
\citep{1992A&A...260...33K,1998ApJ...498L.111D,2021ApJ...911...19P} was adopted, with viewing angles in the range $\theta=20^{\circ}-65^{\circ}$,
as discussed in Sect. \ref{sec:Results}. For these parameters we computed the Doppler factor and obtain $\delta\sim\Dfrange$.
For the optical thin spectral index we estimated $\alpha_{\rm thin}$ based on the flux densities in the radio \citep{1980MNRAS.190..903L} and ultraviolet \citep{2015ApJS..220....6B} bands in the time-averaged spectral energy distribution (SED), obtaining 
$\alpha_{\rm thin} \sim -0.77$.
This value agrees well with other spectral index measurements of the more extended jet emission 
in \C\ seen at longer wavelengths (e.g. \citealt{2000ApJ...530..233W}), as well as with our own results for the optically thin jet region on $\sim$0.5\,mas scales. (Fig. \ref{fig:SpI}). 
\end{appendix}

\end{document}